\documentclass[aps,twocolumn,showpacs,preprintnumbers,amsmath,amssymb]{revtex4}
\usepackage{graphicx}

\begin{document}

\title{Control of atomic decay rates via manipulation of
reservoir mode frequencies }

\author{I.E. Linington} 
\author{B.M. Garraway} 


\begin{abstract} 

  We analyse the problem of a two-level atom interacting with a
  time-dependent dissipative environment modelled by a bath of
  reservoir modes. In the model of this paper the principal
  features of the reservoir structure remain constant in time,
  but the microscopic structure does not. In the context of an
  atom in a leaky cavity this corresponds to a fixed cavity and a
  time-dependent external bath. In this situation we show that by
  chirping the reservoir modes sufficiently fast it is possible
  to inhibit, or dramatically enhance the decay of the atomic
  system, even though the gross reservoir structure is fixed.
  Thus it is possible to extract energy from a cavity-atom system
  faster than the empty cavity rate.  Similar, but less dramatic
  effects are possible for moderate chirps where partial trapping
  of atomic population is also possible.

\end{abstract} 

\date{\today}

\maketitle

\section{Introduction}
\label{sec:intro}

The decay of an excited atomic state is manifestly affected by
its environment. This has been evident since studies began on the
effect of cavities and conducting plates on spontaneous emission
\cite{Purcell46,Kleppner81,Goy83,Jhe86}. The boundary conditions
of nearby surfaces affect the electromagnetic mode density and,
in the simplest treatment, the application of Fermi's golden rule
shows that the spontaneous emission rate is modified. However,
with the development of high-Q cavities in the microwave
\cite{Meschede85} and optical regime \cite{Thompson92} it was
possible to have an oscillatory exchange of energy between a
single atom and a single cavity mode of electromagnetic (EM)
radiation. In this regime a single quantum of EM energy may be
involved and the atom may become entangled with the cavity. It
is often desirable to control quantum systems of this kind, for
example, to try and make a specific state of the cavity field
(see e.g.\ \cite{Walther99}), or to make entanglement, or in more
complex systems, to make photons on demand \cite{pod}. However,
the effect of the environment is always there as decoherence, or
dissipation, and it almost always acts to degrade the quantum
phenomena in which we are interested.

In this paper we will look at a model example of what we might do
in the environment to try and regain control. We will take the
simplest quantum optical system of a two-level atom coupled to a
leaky cavity. In our idealised system the cavity leaks through
lossless dielectric mirrors, although there well may be other
systems that fit the model such as coupled microwave cavities
\cite{Raimond97} or engineered reservoirs for ion traps
\cite{x2}. For the optical cavity system, there are a number of
well established types of quantum optical environment. For
example, the thermal bath contains energy which ultimately leads
to thermalisation of the atom and cavity. Rigged reservoirs
\cite{Dupertuis87} have bath states with coherence which add gain
and noise to the field state. The broad-band squeezed vacuum
\cite{SquReview} has correlations between different frequency
bath states which in general has both the properties of a thermal
bath and the presence of coherence. All of these environments
would eventually excite an initially unexcited atom in the cavity.
The environment we will look at is one in which the bath, or
reservoir modes, change their frequency as a function of time.
This kind of environment would not excite an initially unexcited
atom, although the dynamics of an initially excited atom would be
affected. In general, we would expect that reservoir modes do
have time-dependent properties, since they are themselves
embedded in an uncontrolled environment. However, here we will
suppose that we can at least control their frequency for a short
period of time. The reservoir will be fully specified if the
frequency and the strength of the coupling to the atom are known
for each mode: this amounts to specifying the reservoir mode
structure.

If we modulate all the reservoir mode frequencies by the same
amount, keeping the coupling of each mode to the atom the same,
the overall effect is the same as moving one of the cavity
mirrors, which has been studied for the case of an oscillating
mirror \cite{Law95}. However, this case is also equivalent to a
time-dependent detuning of the reservoir from the atomic
transition frequency. This situation has been discussed in Ref.\
\cite{janowicz2000} where it was found that it was possible to
inhibit the decay of the atom for particular modulation
frequencies. Modulation of the phase of the atom-cavity coupling
constant was also analysed by Agarwal \cite{Agarwal99} who showed
that, for a low-Q cavity situation, the decoherence could be
reduced with fast modulation. Temporal variation of both the
phase and weight of the coupling strengths between an atom and a
continuum, with variation of the atomic detuning was studied in
Ref.\ \cite{kofman2001}. In this case the bath modes had to be
modulated identically and the time-dependence had to be a weak
perturbation. However, it was shown that both enhancement and
suppression of the decay rate relative to the golden rule limit
is possible.

\begin{figure}[]
\includegraphics{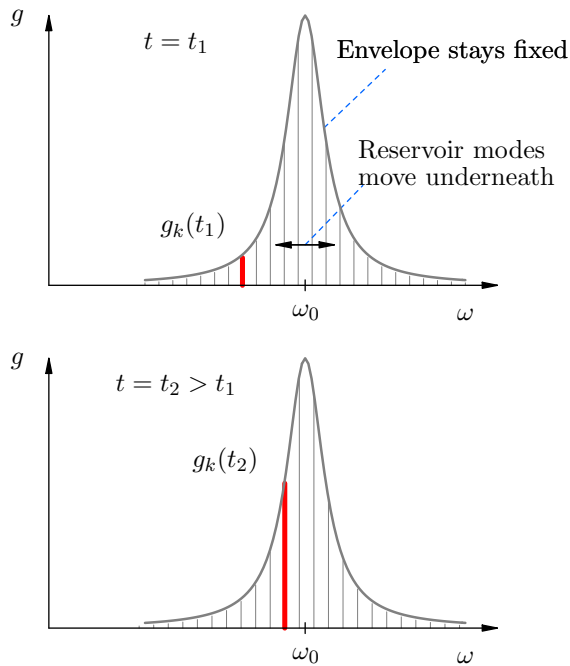}
\caption{ Reservoir mode couplings $g_k(t)$ (see equation
  \ref{Hamiltonian}) at two different times $t_1$ and $t_2$.
  The envelope of the couplings remains the same as the reservoir
  mode frequencies vary in time. Thus the coupling of the atom
  to an individual mode changes in time. The thick line
  indicates the same mode $k$ at two different times, $t_{1}$ and
  $t_{2}$. Since at time $t_{2}$, this mode has moved to a
  different frequency under the static envelope, the coupling has
  changed accordingly.}
\label{modesFig} 
\end{figure}

In this paper we aim to investigate a new and subtle case, where
the mode frequencies and coupling constants are manipulated, but
in such a way as to leave all macroscopic properties of the
reservoir unaffected. The unaffected properties will be the
overall spectral width and resonance frequency of the reservoir
structure. To achieve this the bath modes both increase their
frequency and change their coupling strength as shown in figure
\ref{modesFig}. Passing through the cavity resonance from below,
the coupling strength of a mode to the atom first rises and then
falls as the frequency is increased. The physical realisation
could be a double cavity: one cavity couples to an atom and leaks
to a second cavity which can have its end wall moved. We will
discuss this realisation further in section \ref{conclusions}.

Without further analysis this situation would present us with a
dilemma. The macroscopic structure of the atom-reservoir
interaction is unchanged in time, but the individual bath modes
only interact with the atom for a finite time. It may not be
immediately clear that this situation can dramatically affect the
atomic decay: at any point in time the atom has the same coupling
to some bath modes. However, because the bath modes change, the
history of the atom-bath interaction affects the time development
of their population. We will find that in this case both enhanced
and inhibited decay are possible.

The remainder of this paper is organised as follows. In section
\ref{model} we describe our model for the problem and derive the
equations governing the dynamics of the atomic state populations.
Section \ref{static_reservoir} reviews the standard problem of a
two-level atom interacting with a high-Q cavity in the strong and
weak coupling regimes. However, the section pays particular
attention to the evolution of the quantum state of the reservoir
degrees of freedom, which are often traced over in similar
calculations. Section \ref{results} summarises the main results
of the investigation in the regimes of strong and weak coupling.
Finally, in section \ref{conclusions}, we estimate the size of
the effects that might be observed and summarise the findings of
the paper.

\section{Dynamic Reservoir Model}
\label{model}
We consider a two-level atom, with states denoted
$\vert{}0\rangle$ and $\vert{}1\rangle$, coupled to a reservoir
of electromagnetic radiation modes at zero temperature. The
reservoir modes, labelled by an index $k$, have
\emph{time-dependent} frequencies $\omega_k(t)$, and raising and
lowering operators $\hat{b}_{k}^{\dagger}$ and $\hat{b}_{k}$. The
Hamiltonian for the evolution of the composite atom and reservoir
system is given in the rotating wave approximation by ($\hbar=1$)
\begin{eqnarray}
\hat{H}(t) & = &
\omega_{0}\hat{\sigma}^{+}\hat{\sigma}^{-}+\sum_{k}\omega_k(t)\left(\hat{b}_{k}^{\dagger}\hat{b}_{k}+1/2\right)
\nonumber \\ & &
+\sum_{k}g_k(t)\left(\hat{\sigma}^{-}\hat{b}_{k}^{\dagger}+\hat{b}_{k}\hat{\sigma}^{+}\right),
\label{Hamiltonian}
\end{eqnarray}
where the atomic transition frequency is $\omega_{0}$, and
$\hat{\sigma}^{+}=\vert{}1\rangle\langle{}0\vert$ and
$\hat{\sigma}^{-}=\vert{}0\rangle\langle{}1\vert$ are the
atomic raising and lowering operators respectively. The electric
dipole coupling of the $k^{th}$ mode of the electromagnetic
radiation field to the atomic transition
$\vert{}0\rangle\leftrightarrow\vert{}1\rangle$ is denoted
by $g_k(t)$. Without loss of generality, the $g_k(t)$ are chosen
to be real. For simplicity, we will later choose the spectral
profile of the coupling constants to be peaked at the frequency
of the atomic transition. While the reservoir modes are
re-distributed, we will suppose that this envelope remains fixed.
Therefore, the magnitude of the coupling constants $g_k(t)$ must
vary in time to ensure this (see figure \ref{modesFig}). We will
also assume that whilst the frequency change of the reservoir
modes may be very rapid from the point of view of the atom, it
will not be so fast that it could create photons in the cavity by
itself. In this sense the cavity field will adiabatically follow
the motion of its end-mirror (see, e.g., Refs.\
\cite{janowicz1998,dodonov1996}).

To develop equations for the state vector we work in an
interaction picture and so we may write the time-evolution
operator as
\begin{eqnarray}
\hat{U}(t,0) & = & \exp\bigg(-i\int_{0}^{t}\sum_{k}\omega_k(\tau)\left(\hat{b}_{k}^{\dagger}\hat{b}_{k}+1/2\right)d\tau\bigg)
\nonumber \\
& &
\times\exp\bigg(-i\omega_{0}t\hat{\sigma}^{+}\hat{\sigma}^{-}\bigg)\hat{U}_{I}(t,0).
\end{eqnarray}
The effect of the atom-reservoir interaction is contained in the
term $\hat{U}_{I}(t,0)$, which satisfies the Schr\"odinger equation
\begin{align}
i\frac{d\hat{U}_{I}(t,0)}{dt}=\hat{H}_{I}(t)\hat{U}_{I}(t,0),
\label{UI_def}
\end{align}
and the interaction picture Hamiltonian $\hat{H}_{I}(t)$ is given by
\begin{align}
\hat{H}_{I}(t) = &
\sum_{k}\bigg[g_k(t)\hat{\sigma}^{-}\hat{b}_{k}^{\dagger}\exp\left(i\int_{0}^{t}
\big[\omega_k(\tau) - \omega_{0}\big]
d\tau\right) \nonumber \\ 
&  +g_k(t)\hat{\sigma}^{+}\hat{b}_{k}\exp\left(-i\int_{0}^{t}\big[\omega_k(\tau') - \omega_{0}\big]d\tau'\right)\bigg].
\nonumber \\
\label{H_int}
\end{align}
The state vector is expanded in terms of the eigenstates of the
excitation number operator,
$\hat{N}=\hat{\sigma}^{+}\hat{\sigma}^{-}+\sum_{k}\hat{b}_{k}^{\dagger}\hat{b}_{k}$.
Since $\hat{N}$ commutes with the Hamiltonian (\ref{H_int}), the
number of quanta is a constant of the motion. For initial states
we assume the system is prepared with the atom in its excited
state, and the reservoir is initially empty, i.e.\
$\vert\psi(0)\rangle=\vert1\rangle\otimes\vert\ldots{}0\ldots\rangle$.
Then the state vector at time $t$ may be written as
\begin{align}
\left\vert\psi(t)\right\rangle = & \hat{U}_{I}(t,0)\left\vert\psi(0)\right\rangle \nonumber \\
 =&
c_{a}(t)\left\vert1\right\rangle\otimes\left\vert\ldots0\ldots\right\rangle \label{state_vector}
\\ & +
\sum_{k}c_{k}(t)\left\vert0\right\rangle\otimes\left\vert\ldots1_{k}\ldots\right\rangle
\nonumber
\end{align}
where $c_{a}(t)$ is the amplitude for the atom to be found in its
excited state, and $c_{k}(t)$ is the amplitude for the $k^{th}$
mode of the reservoir to contain one excitation. Elimination of
$\hat{H}_{I}(t)$ and $\hat{U}_{I}(t)$ from equations
(\ref{UI_def}), (\ref{H_int}) and (\ref{state_vector}) leads to
the following coupled differential equations for the amplitudes
$c_{a}(t) $ and $c_{k}(t)$:
\begin{align}
i\frac{\partial{}c_{a}(t)}{\partial{}t} = & \sum_{k}g_{k}\left(t\right)\exp\left(-i\int_{0}^{t}\big[\omega_k(\tau) - \omega_{0}\big]d\tau\right)c_{k}(t)
\label{chirp_dynamics1} \\
i\frac{\partial{}c_{k}(t)}{\partial{}t} = & g_{k}\left(t\right)\exp\left(i\int_{0}^{t}\big[\omega_k(\tau) - \omega_{0}\big]d\tau\right)c_{a}(t).
\label{chirp_dynamics2}
\end{align}
We can now eliminate the amplitudes $c_{k}(t)$ by integrating
equation (\ref{chirp_dynamics2}) and substituting the result into
equation (\ref{chirp_dynamics1}). Thus we obtain the
integro-differential equation for $c_{a}(t)$, i.e.
\begin{align}
\frac{\partial{}c_{a}(t)}{\partial{}t} = &
-\int_{0}^{t}dt'\sum_{k}\; g_k(t)g_k(t') \nonumber \\ & 
\times\exp\left(-i\int_{0}^{t}\big[\omega_k(\tau) - \omega_{0}\big]d\tau\right)
\label{ca_integro_diff1} \nonumber \\
&
\times\exp\left(i\int_{0}^{t'}\big[\omega_k(\tau') - \omega_{0}\big]d\tau'\right)c_{a}(t').
\end{align}
For much of the following analysis we will work in the limit of a
continuum of reservoir modes so that the sum in
(\ref{ca_integro_diff1}) is replaced by an integral over a
continuous variable $\omega_{}$. Thus we make the
identifications
\begin{align}
  \omega_k(t) \rightarrow& \omega_R(\omega,t)\\
  g_k(t) \rightarrow& g(\omega,t)
\label{eq:k_identifications}
\end{align}
where $\omega_{}$ refers to the \emph{initial} frequencies of the
field modes. The function $\omega_R(\omega,t)$ represents, at
time $t$, the frequency of a bath mode which had frequency
$\omega$ at $t=0$. Hence, we will have
$\omega_R(\omega,0)=\omega$. We also introduce the density of
states $\rho(\omega)$ at time $t=0$ and then equation
(\ref{ca_integro_diff1}) is replaced by
\begin{align}
\frac{\partial{}c_{a}(t)}{\partial{}t} \approx &
-\int_{0}^{t}dt'\int_{-\infty}^{\infty}d\omega_{}\;
\rho(\omega)g(\omega,t)g(\omega,t')
\nonumber \\ & 
\times\exp\left(-i\int_{0}^{t}\big[\omega_R(\omega,\tau) - \omega_{0}\big]d\tau\right)
\label{caid}\\ &
\times\exp\left(i\int_{0}^{t'}\big[\omega_R(\omega,\tau') - \omega_{0}\big]d\tau'\right)
c_{a}(t'). \nonumber
\end{align}
In addition, in equation (\ref{caid}), we have extended the lower
frequency limit from $0$ to $-\infty$. However, for optical
transitions, ($\omega_{0}\sim10^{15}s^{-1}$), the effects
of this approximation are similar to others already made (for
example the rotating-wave approximation, or the assumption of a
two-level atom), and may be neglected.

Later, we will also examine an idealised time-dependent bath
spectrum, which is in fact just the frequency dependent
population of the reservoir modes. We define this spectrum by
considering those modes $k_\Delta$ ($k_\Delta \in \{k\}$) which
have frequencies that lie within $\Delta\omega$ of a given
frequency $\omega$ at a time $t$. That is, we let
\begin{equation}
  S(\omega,t) \approx \frac{1}{\Delta\omega} \sum_{k_\Delta}
  |c_{k_\Delta}(t)|^2
  .
\label{eq:spectrum.approx}
\end{equation}
In the continuum limit the number of modes $k_\Delta$ in the sum
is approximately $\Delta\omega \rho(\omega)$, and since the
$c_{k}$ are expected to vary smoothly with $k$ in this limit, we
finally let
\begin{equation}
   S(\omega,t)  \longrightarrow  \rho(\omega)
   |c_{k_\Delta}(t)|^2
   ,
\label{eq:spectrum.def}
\end{equation}
which applies to a representative $k_\Delta$. Equation
(\ref{eq:spectrum.def}) will serve as an operational definition
of the bath spectrum.

We will now specify a model reservoir structure to be used in
equation (\ref{caid}). For a simple cavity model, we can assume
that the single time reservoir structure function
$\rho(\omega)\vert{}g(\omega,t)\vert^{2}$ is a Lorentzian, with
width $\gamma$, centred on the atomic transition frequency
$\omega_{0}$, i.e.
\begin{align}
\rho(\omega)\vert{}g(\omega,t)\vert^2 =  
\frac{D^2\gamma/\pi}{\gamma^2+(\omega_R(\omega,t) - \omega_{0})^2}.
\label{rho_g_squared}
\end{align}
This describes the time-dependent coupling of a bath mode,
initially at $\omega$. However, it also describes the
instantaneous coupling to the bath of modes at time $t$. Since
the structure has no time dependence, apart from the time
dependence of $\omega_R$, the basic reservoir structure is
constant in time, in this model. This is also clear from the
weight $D$ of the Lorentzian, which is defined through the
relation
\begin{equation}
\int_{-\infty}^{\infty}\rho(\omega)\vert{}g(\omega,t)\vert^{2}
d\omega_{} = D^{2}
.
\end{equation}
We can see that this integral is independent of time for the
chosen form (\ref{rho_g_squared}). In the static case, where
$\rho(\omega)\vert{}g(\omega,t)\vert^2$ is not a function of
time, the Lorentzian (\ref{rho_g_squared}) is a common choice for
the reservoir structure \cite{barnett1997,lang1973} and the
resulting dynamics have been well explored (see, for example,
\cite{garraway1997, lambropoulos2000}). In the weak coupling
limit, the Lorentzian reservoir structure ensures exponential
decay of atomic population. Of course, equation (\ref{caid})
does not feature $\rho(\omega)\vert{}g(\omega,t)\vert^{2}$, but
the two-time product $\rho(\omega)g(\omega,t)g(\omega,t')$.
However, as already mentioned, $g(\omega,t)$ may be chosen to be
real. Therefore the two-time product follows immediately as
\begin{align}
\rho(\omega)g(\omega,t)g(\omega,t') = & \frac{D^2\gamma}{\pi\sqrt{\left(\gamma^2+(\omega_R(\omega,t) - \omega_{0})^2\right)}}
\nonumber \\
&\times\frac{1}{\sqrt{\left(\gamma^2+(\omega_R(\omega,t') - \omega_{0})^2\right)}}.
\label{rho_g(t)_g(t')}
\end{align}
To proceed with the analysis, a specific form for the time
dependence of the modes must also be chosen. In this paper we
choose a simple case where all the bath-mode frequencies increase
linearly at the same rate, i.e.\ a linear chirp, for which
\begin{equation}
\omega_R(\omega,t) = \omega_{} + \chi{}t.
\label{frequency_definition}
\end{equation}
Some motivation for this will be given in section
\ref{conclusions}. If we now substitute
(\ref{frequency_definition}) and (\ref{rho_g(t)_g(t')}) into
(\ref{caid}) we find
\begin{align}
\frac{\partial{}c_{a}(t)}{\partial{}t} = &
-\frac{D^2\gamma}{\pi}\int_{0}^{t}dt'
\int_{-\infty}^{\infty}d\omega_{}\frac{\exp\left[-i\left(\omega_{}  -
      \omega_{0}\right)t- i\chi t^2 /2 \right]}{\sqrt{\left(\gamma^2+\left(\omega_{} - \omega_{0}+ \chi{}t\right)^2\right)}}
\nonumber \\ & 
 \hspace{10mm}\times\frac{\exp\left[i\left(\omega_{} - \omega_{0}\right)t'+i\chi t'^2/2\right]}{\sqrt{\left(\gamma^2+\left(\omega_{} - \omega_{0}+\chi{}t'\right)^2\right)}}\;c_{a}(t') \nonumber \\
= & -\int_{0}^{t}dt'K(t,t')c_{a}(t'),
\label{ca_integro_diff2}
\end{align}
i.e.\ an integro-differential equation with the kernel given by
\begin{align}
K(t,t') = &
\frac{D^2\gamma}{\pi}
\int_{-\infty}^{\infty}d\omega_{}\frac{\exp\left[-i\left(\omega_{}  -
      \omega_{0}\right)t- i\chi t^2 /2\right]}{\sqrt{\left(\gamma^2+\left(\omega_{} - \omega_{0}+ \chi{}t\right)^2\right)}}
\nonumber \\ & 
\hspace{8mm}\times
 \frac{\exp\left[i\left(\omega_{} -
       \omega_{0}\right)t'+ i\chi t'^2 /2\right]}{\sqrt{\left(\gamma^2+\left(\omega_{} - \omega_{0}+\chi{}t'\right)^2\right)}} .
\label{kernel}
\end{align}
We also note that if we change variables to $\tau=t-t'$, and
$\Delta = \omega-\omega_0 + \chi (t+t')/2$, equation
(\ref{ca_integro_diff2}) can be expressed as
\begin{widetext}
\begin{align}
  \frac{\partial{}c_{a}(t)}{\partial{}t} = &
-\frac{D^2\gamma}{\pi}\int_{0}^{t}d\tau 
\int_{-\infty}^{\infty}d\Delta
\frac{ e^{-i\Delta\tau}  c_a(t-\tau)
}{
\sqrt{
\left[ \gamma^2 + (\Delta + \chi\tau/2)^2 \right]
\left[ \gamma^2 + (\Delta - \chi\tau/2)^2 \right]
}}
  .
  \label{eq:integro_diff3}
\end{align}
\end{widetext}
\section{Evolution of a Static Reservoir}
\label{static_reservoir}
We are investigating the effects of manipulating a reservoir of
field modes which will interact with an atom. That is, the
reservoir changes in time. However, it will be helpful to discuss
first the behaviour of the static problem. Here, none of the
reservoir mode frequencies vary with time ($\chi=0$), so equation
(\ref{ca_integro_diff2}) simplifies to
\begin{align}
\frac{\partial{}c_{a}(t)}{\partial{}t} = & -\frac{D^2\gamma}{\pi}\int_{0}^{t}dt'\int_{-\infty}^{\infty}d\omega_{}\frac{e^{-i\left(\omega_{} - \omega_{0}\right)\left(t-t'\right)}}{\gamma^2+\left(\omega_{} - \omega_{0}\right)^2}c_{a}(t') 
\nonumber \\
= & -D^{2}\int_{0}^{t}e^{-\gamma\left(t-t'\right)}c_{a}(t')dt'
\nonumber \\
= & -\int_{0}^{t}K_{s}(t,t')c_{a}(t')dt'.
\label{ca_integro_diff_static}
\end{align}
Equation (\ref{ca_integro_diff_static}) can, for example, be
solved for the dynamics of the atom by means of Laplace
transforms. We may also use the method of \emph{pseudomodes}
\cite{garraway1997}. For this system a pseudomode amplitude
$\mathcal{B}(t)$ is defined as
\begin{align}
\mathcal{B}(t) = & -iDe^{-\gamma{}t}\int_{0}^{t}e^{\gamma{}t'}c_{a}(t')dt',
\end{align}
and will represent the dynamics of the field. The dynamics of
$c_{a}(t)$ are then contained in the following two equations
which can be exactly solved and represent equation
(\ref{ca_integro_diff_static}) \cite{garraway1997}:
\begin{align}
\frac{\partial{}c_{a}(t)}{\partial{}t} = & -iD\mathcal{B}(t) 
\label{pseudomode_1} \\
\frac{\partial{}\mathcal{B}(t)}{\partial{}t} = & -iDc_{a}(t) - \gamma{}\mathcal{B}(t).
\label{pseudomode_2}
\end{align}
If the frequency scale associated with the reversible dynamics of
the atom-field coupling $D$ exceeds that for irreversible decay
$\gamma$, then there will be a resonant exchange of energy
between atom and cavity (Rabi oscillations) as a photon emitted
by the atom is repeatedly emitted and re-absorbed. This is the
regime of \emph{strong coupling}. We note that in the weak
coupling limit the decay of the \emph{cavity field} is seen to be
given by $\gamma$.  To improve the clarity of later discussions,
we will now analyse the dynamics separately in the
strong-coupling and weak-coupling regimes.

\subsection{Strong Coupling}
\label{static_reservoir_strong}
In the regime of strong coupling to the static reservoir
($D\gg\gamma$), the exact solution to equations
(\ref{pseudomode_1},\ref{pseudomode_2}) can be written as
\begin{equation}
c_{a}(t)  =  e^{-\gamma t/2}   \left[\cos\left(\Omega
    t/2\right)+\frac{\gamma}{\Omega}\sin\left(\Omega t/2 \right)\right],
\label{exact_ca_static}
\end{equation}
where \cite{lambropoulos2000}
\begin{align} 
\Omega = \sqrt{4D^2-\gamma^2}.
\label{Omega_definition}
\end{align}
The population of the excited atomic state then undergoes Rabi
oscillations at angular frequency $\Omega$ ($\approx{}2D$), under
an envelope decaying at rate $\gamma$. The atomic decay at a rate
$\gamma$ is a feature of strong coupling to the cavity,
$D>\gamma/2$ in equation (\ref{Omega_definition}), and the chosen
reservoir structure (\ref{rho_g_squared}).

\begin{figure}[]
\includegraphics[width=0.8\columnwidth]{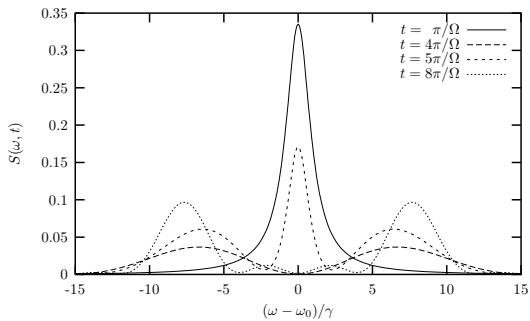}
\caption{ Occupation of reservoir field-modes $S(\omega,t)$ at
  four different times. The times are chosen so that the atomic
  excited state probability, $\vert{}c_{a}(t)\vert^{2}$ is either
  zero ($t= \pi/\Omega, 5\pi/\Omega$), or at a local maximum
  ($t=4\pi/\Omega, 8\pi/\Omega$). The reservoir is static
  ($\chi=0$) and modes in the range $\omega_{0}-\Omega/2$ to
  $\omega_{0}+\Omega/2$ are occupied over the four Rabi cycles
  shown here. The parameter $D=8\gamma$ (strong coupling).  The
  calculation is based on the substitution of equation
  (\ref{exact_ca_static}) into (\ref{chirp_dynamics2}).}
\label{intermediate_chirp_fig}
\end{figure}

For later comparison we will determine the bath spectrum
$S(\omega,t)$ defined in equation (\ref{eq:spectrum.def}). In
principle this can be found by inserting the solution
(\ref{exact_ca_static}) into equation (\ref{chirp_dynamics2}) and
integrating. The resulting expression for $S(\omega,t)$ is very
complicated and here we will instead derive this spectrum in the
strong coupling limit where $D\gg\gamma$. Thus we take the
solution (\ref{exact_ca_static}) to be approximated by
\begin{equation}
c_{a}(t) \approx e^{-\gamma{}t/2}\cos(\Omega t /2) .
\end{equation}
Substituting the expression into (\ref{chirp_dynamics2}) and
integrating (using $c_{k}(0)=0$) gives
\begin{align}
c_{k}(t) & = &
\frac{g_k}{2}
\left\{\frac{\exp\left[i(\omega_k-\omega_0-\Omega/2+ i\gamma/2)t\right]-1}{\omega_k-\omega_0-\Omega/2+i\gamma/2}
\right.
\nonumber \\
& &
\left.
  +\frac{\exp\left[i(\omega_k-\omega_0+\Omega/2+
    i\gamma/2 )t\right]-1}{\omega_k-\omega_0+
  \Omega/2+ i\gamma/2 } \right\}
.
\label{eq:ck_static}
\end{align}
We now substitute (\ref{eq:ck_static}) into the definition for
the spectrum, equation (\ref{eq:spectrum.def}), and obtain an
expression for $S(\omega,t)$ which simplifies in this strong
coupling limit. This is because $S(\omega,t)$ is small away from
either $\omega_{}\approx\omega_{0}$ or
$\omega_{}\approx\omega_{0}\pm \Omega/2 $. As a result, if we
also use the replacement $\omega_k \rightarrow \omega$ and the
reservoir structure function (\ref{rho_g_squared}), the spectrum
can be simplified to:
\begin{widetext}
\begin{align}
S(\omega,t)
\approx & 
\frac{\gamma/2}{2\pi\left[(\omega_{}-\omega_{0}- \Omega/2 )^2+\left( \gamma/2 \right)^2\right]}\left[1+e^{-\gamma{}t}-2e^{-\gamma{}t/2}\cos\left(\left(\omega_{}-\omega_{0}- \Omega/2 \right)t\right)\right]
\nonumber \\
& + \frac{\gamma/2}{2\pi\left[(\omega_{}-\omega_{0}+ \Omega/2 )^2+\left( \gamma/2 \right)^2\right]}\left[1+e^{-\gamma{}t}-2e^{-\gamma{}t/2}\cos\left(\left(\omega_{}-\omega_{0}+ \Omega/2 \right)t\right)\right]
\nonumber \\
& +
\frac{\gamma}{\pi\left[\left(\omega_{}-\omega_{0}\right)^2+\gamma^2\right]}e^{-\gamma{}t}\sin^2\left(
  \Omega t/2\right)
.
\label{occupations_2} 
\end{align}
In the long time limit the atom has completely decayed and the
initial energy is distributed in the bath as
\begin{equation}
S(\omega,t\rightarrow\infty) =
\frac{1}{2\pi}\left[
\frac{\gamma/2}{(\omega_{}-\omega_{0}- \Omega/2 )^2+\left( \gamma/2 \right)^2}+\frac{\gamma/2}{(\omega_{}-\omega_{0}+ \Omega/2 )^2+\left( \gamma/2 \right)^2}
\right].
\end{equation}
(See also Ref.\ \cite{Carmichael89}.)
\end{widetext}
It is seen from equation (\ref{occupations_2}) that the atom
interacts with modes in the approximate range
$\omega_{0}-\Omega/2$ to $\omega_{0}+\Omega/2$ (in the sense that
only modes in this range become significantly changed via
interaction with the atom). As the interaction progresses the
reservoir mode occupation probabilities develop away from the
atomic frequency $\omega_{0}$ at $\omega_{0}\pm \Omega/2 $ (the
vacuum Rabi-splitting, \cite{sanchez1983,boca2004}). For the
\emph{intermediate coupling case}, where the tails of the
Lorentzians in equation (\ref{occupations_2}) overlap, the
population appears to move outwards (figure
\ref{intermediate_chirp_fig}). However, equation
(\ref{occupations_2}) is still approximately valid. For this
intermediate coupling the system does undergo Rabi oscillations,
($D>\gamma/2$), without the strong coupling condition
$D\gg\gamma$ being satisfied. In this same regime we find that
chirping the reservoir mode frequencies has interesting effects
as we will see in the next section. This is because the occupied
modes at frequencies less than $\omega_{0}$ may be brought back
into resonance with the atom through the reservoir frequency
manipulation and thus affect $\vert{}c_{a}(t)\vert^2$ via
equation (\ref{chirp_dynamics1}).

\subsection{Weak Coupling}
\label{static_reservoir_weak}

For reference later we note below the behaviour of the atomic
dynamics in the weak coupling regime ($D\ll\gamma$). The exact
solution to equations (\ref{pseudomode_1},\ref{pseudomode_2})
shows damping at two different rates given approximately as
$D^2/\gamma$ and $\gamma$. The damping at rate $\gamma$ only has
a small effect at short times. An alternative approach notes that
the kernel $K_{s}(t,t')$ in equation
(\ref{ca_integro_diff_static}) is negligible away from
$t'\approx{}t$. Then the coefficient $c_{a}(t')$ can only
contribute to the integral at $t'\approx{}t$, so we may make the
\emph{Markov approximation}, and equation
(\ref{ca_integro_diff_static}) simplifies to
\begin{align}
\frac{\partial{}c_{a}(t)}{\partial{}t}  \approx & - c_{a}(t)\int_{0}^{t}K_{s}(t,t')dt'
\nonumber \\
 \approx & -\frac{\Gamma(t)}{2}\;c_{a}(t),
\label{markov1}
\end{align}
where,
\begin{align}
\Gamma(t)  = &
\frac{2D^2}{\gamma}\left(1-e^{-\gamma{}t}\right).
\label{Weak:TDPgamma}
\end{align}
This time-dependent decay rate gives atomic dynamics which are
correct to second order in $(D/\gamma)^2$ and approximates the
short-time dynamics as well as the long time behaviour. For the
latter, we see that the term $e^{-\gamma{}t}$ in $\Gamma(t)$
decreases rapidly over the decay-time of the atom, so we may then
let
\begin{equation}
\Gamma(t)\approx\frac{2D^2}{\gamma},
\label{markov_const}
\end{equation}
resulting in exponential time-dependence for the atom: $c_{a}(t)
\approx \exp( -D^2 t / \gamma )$. The atom and reservoir become
entangled and the amplitude $c_{a}(t)$ decays at a rate which is
always less than the strong coupling case, equation
(\ref{exact_ca_static}).

\section{Results for a time-dependent reservoir}
\label{results}

Exact, analytic solution of equation (\ref{ca_integro_diff2}) has
not proved possible. Square roots in the denominator introduce
branch cuts in the complex angular frequency plane, and this
creates difficulties in more than one approach. Therefore the
solution to this problem must be numerical. However, in certain
limits, an approximate analytical solution is possible, and the
results are presented in sections \ref{results_strong_high},
\ref{results_strong_low}, and \ref{results_weak}.

Where a numerical solution for the amplitudes $c_{a}(t)$ and
$c_{k}(t)$ has been required, it was obtained by integrating
(\ref{chirp_dynamics1}) and (\ref{chirp_dynamics2}) for a
discrete micro-bath of field modes, with the form
(\ref{rho_g_squared}) for the coupling constants, and linear
chirp (\ref{frequency_definition}) for the time-dependence of the
reservoir mode frequencies. The numerical method used was a
fourth-order Runge-Kutta stepwise integration, with adaptive step
size. A suitable density of states was chosen and the results
were checked to make sure the dynamics of
$\vert{}c_{a}(t)\vert^2$ and $\vert{}c_{k}(t)\vert^2$ were
insensitive to this choice. (Typically about ten modes within one
half-width $\gamma$ of the reservoir structure function
(\ref{rho_g_squared}) was sufficient.) Results of this simulation
were tested in the static ($\chi=0$) case against the exact
solution afforded by the method of pseudomodes (equation
(\ref{exact_ca_static}) and reference \cite{garraway1997}), and
for very high chirp rates against analytical predictions as
discussed later in this paper (equation (\ref{Markovian_decay})).

\subsection{Strong Coupling}
\label{results_strong}
\begin{figure}[]
\includegraphics{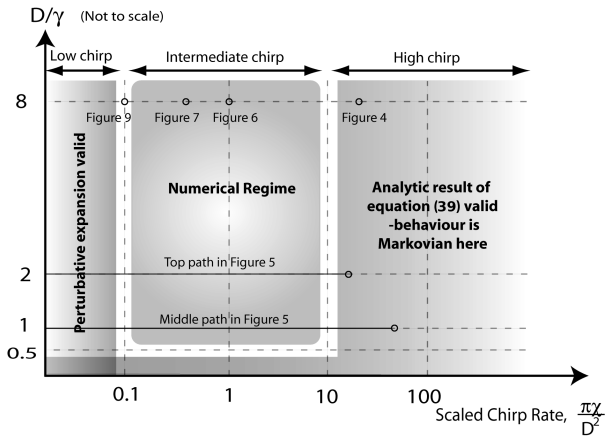}
\caption{Parameter space portrait highlighting the different
  character of solutions to equation (\ref{ca_integro_diff2}) in
  different regions. The specific cases in figures
  \ref{high_chirp_decay_fig1}-\ref{high_Q_low_chirp_fig} are
  indicated. The analytic result of equation
  (\ref{final_Gamma_infinity}) holds in the weak-coupling regime
  $(D<\gamma/2)$, and also in the `high-chirp' regime
  $(\xi\gg1)$, irrespective of $D/\gamma$. }
\label{landscape_fig}
\end{figure}

The effects of reservoir chirp are examined in three regimes:
low, intermediate and high chirp-rates. The classification
depends on the rate of change of the mode frequencies $\chi$
\cite{note1}, the time-scale of the atomic dynamics, and also the
width of the occupied mode spectrum. As discussed in section
\ref{static_reservoir}, even in the absence of frequency
modulation, the time-scale of the atomic dynamics depends on
ratio of the weight $D$ to the width $\gamma$ of the coupling
spectrum. For strongly coupled systems ($D\gg\gamma$), the
relevant time-scale is the Rabi period $2\pi/\Omega$, and since
modes in the frequency range $\omega_{0}- \Omega/2 $ to
$\omega_{0}+ \Omega/2 $ interact significantly with the atom, a
relevant parameter is the ratio of the frequency increase that
each mode experiences during one Rabi period to the frequency
range half-splitting $\Omega/2$. Thus we introduce
\begin{equation}
\xi = \frac{4\pi\chi{}}{\Omega^{2}}
\label{eq:def_xi}
\end{equation}
as the parameter we will use to characterise the different types
of behaviour. Then the specification of only two parameters in
the model completely determines the dynamics: i.e.\ $D/\gamma$
(the strength of the atom-cavity coupling) and the scaled chirp
rate $\xi$. Within this plane in parameter space, three
categories of solution are found, as shown in figure
\ref{landscape_fig}. Here `low' chirp corresponds to $\xi\ll 1$,
`high' chirp to $\xi\gg 1$, and all three categories will be
discussed separately below. Note that the definition of $\xi$
only makes sense for strong coupling, $D> \gamma/2 $, as this is
the region in which $\Omega$ is real and non-zero (see equation
(\ref{Omega_definition})). However, in figure
\ref{landscape_fig}, to include both the strong and weak coupling
regimes, we use $\pi\chi/D^{2}$ for the scaled chirp-rate,
instead of $\xi$, above. For most of the plane shown,
$\Omega\approx2D$, and so this difference is minor.

\subsubsection{High Chirp Rates}
\label{results_strong_high}
\begin{figure}[b]
\includegraphics[width=0.8\columnwidth]{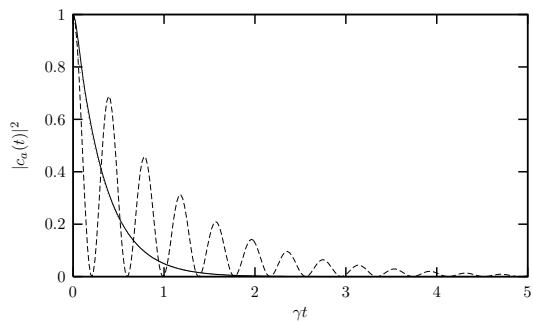}
\caption{Conversion from non-Markovian to Markovian dynamics by
  the introduction of reservoir chirp.  In this example the
  dynamics were non-Markovian (dashed line) when the atom
  interacted with a static reservoir, as manifested by the Rabi
  oscillations. Chirp in the reservoir frequencies resulted in
  Markovian dynamics for the atom, and an exponential decay of
  $\vert{}c_{a}(t)\vert^{2}$ (solid line). The parameters used
  were $D=8\gamma$ and for the chirped case, $\chi=400\gamma^2$,
  which gives $\xi\approx{}20$. Overlaid on the solid line (and
  nearly indistinguishable from it), is the analytic result of
  equations (\ref{final_Gamma_infinity},\ref{Markovian_decay})
  (shown dotted) where $\Gamma_\infty \approx 3\gamma$. }
\label{high_chirp_decay_fig1}
\end{figure}

In the limit $\xi\gg1$, each mode of the reservoir is effectively
only coupled to the atom for a small fraction of one Rabi period.
During this time, the occupation of a mode may increase slightly
through equation (\ref{chirp_dynamics1}), but does not grow large
enough to re-populate the atomic excited state via equation
(\ref{chirp_dynamics2}). Therefore, although the atom-reservoir
coupling is strong, Rabi oscillations are not observed.
Manipulating the reservoir frequencies can change the \emph{type}
of atomic state behaviour from \emph{non-Markovian} to
\emph{Markovian}. A specific example is shown in figure
\ref{high_chirp_decay_fig1}, which lies in the region of high
chirp rate indicated on figure \ref{landscape_fig}. In this case
there is a dramatic difference in the evolution of the atomic
population between the static reservoir (dashed oscillating
curve) and the chirped reservoir (exponentially decaying curve).

\begin{figure}[]  
\includegraphics[width=0.8\columnwidth]{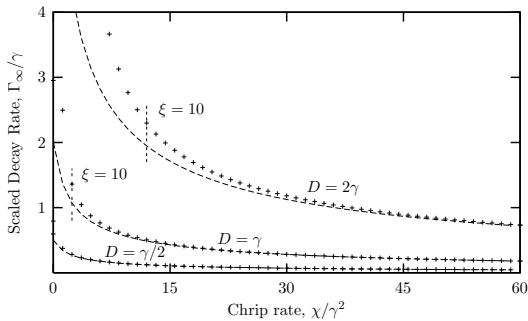}
\caption{Dependence of decay rate $\Gamma_{\infty}$ on the
  chirp-rate $\chi$. The dashed lines show the prediction of
  equation (\ref{final_Gamma_infinity}), whereas the crosses show
  the decay rates extracted from numerical solution (without
  making the Markov approximation) by fitting an exponential
  decay to the data. The parameters used for the three curves
  were $D=\gamma/2, \gamma$, and $2\gamma$. Note that the fit is
  only expected to be good for $\xi\gg1$, since for smaller
  chirp-rates than this, the behaviour of
  $\vert{}c_{a}(t)\vert^2$ is not necessarily an exponential
  decay and may even be oscillatory. Thus for the highest two
  values of $D$ the position of $\xi=10$ is marked. The chirp
  $\chi$ enhances the decay when $\Gamma_{\infty}/\gamma>1$. }
\label{DecayRateFig}
\end{figure}

To investigate this we note that when we chirp the mode
frequencies at a high rate, the two-time product
$\rho(\omega)g(\omega,t)g(\omega,t')$ of equation
(\ref{rho_g(t)_g(t')}) is only non-negligible when $t
\approx{}t'$, i.e.\ the Lorentzians under the square-root
separate when $\chi(t-t')\sim \gamma$. Making the Markov
approximation, the integro-differential equation
(\ref{ca_integro_diff2}) becomes
\begin{align}
\frac{\partial{}c_{a}(t)}{\partial{}t} \approx &
-c_{a}(t)\frac{D^2\gamma}{\pi}\int_{-\infty}^{\infty}d\omega_{}\frac{\exp\left[-i\left(\omega_{}
      -
      \omega_{0}\right)t-\frac{i\chi{}}{2}t^2\right]}{\sqrt{\left(\gamma^2+\left(\omega_{} - \omega_{0} +\chi{}t\right)^2\right)}}
\nonumber \\ &
\qquad
\times\int_{0}^{t}dt'\frac{\exp\left[i\left(\omega_{} - \omega_{0}\right)t'+\frac{i\chi{}}{2}t'^2\right]}{\sqrt{\left(\gamma^2+\left(\omega_{} - \omega_{0}+\chi{}t'\right)^2\right)}}
\nonumber \\  
\approx & -c_{a}(t)\int_{0}^{t}dt'K(t,t')
\nonumber \\
\approx & -\frac{\Gamma(t)}{2}\;c_{a}(t),
\label{ca_high_chirp1}
\end{align}
where we have defined a time-dependent decay rate as
\begin{equation}
  \Gamma(t)= 2 \int_{0}^{t}K(t,t')dt' .
\end{equation}
Time-dependent decay rates similar to $\Gamma(t)$ also arise in
the time-convolutionless projection operator technique \cite{TCL}
in the second order approximation. We already encountered a
similar time-dependent decay rate in the weakly coupled static
reservoir in equation (\ref{Weak:TDPgamma}). The decay rate
$\Gamma(t)$ tends to a constant value as
$t\longrightarrow\infty$, which corresponds to a Markovian limit.
This limit is given by
\begin{widetext}
\begin{align}
\Gamma_{\infty} = & \lim_{t\rightarrow\infty}\bigg\{\Gamma(t)\bigg\}
\nonumber \\
= & \lim_{t\rightarrow\infty}\Bigg\{\frac{2D^2\gamma}{\pi}\int_{0}^{t}\int_{-\infty}^{\infty}\frac{\exp\left(i\left(\omega_{}-\omega_{0} +\chi{}t'/2\right)t'\right)}{\sqrt{\gamma^2+\left(\omega_{}-\omega_{0}+\chi{}t'\right)^2}}
\nonumber \\
& \qquad  \times
\frac{\exp\left(-i\left(\omega_{}-\omega_{0}+\chi{}t/2)\right)t\right)}{\sqrt{\gamma^2+\left(\omega_{}-\omega_{0}+\chi{}t\right)^2}}d\omega_{}dt'\Bigg\}.
\label{Gamma_infinity1} 
\end{align}
To evaluate this we may change to variables $\Delta_d
=\chi(t-t')/2$ and $\Delta_{} = \omega_{}-\omega_{0}+
\chi(t+t')/2$, so that
\begin{align}
\Gamma_{\infty} = & \frac{4D^2\gamma}{\pi\chi}\int_{0}^{\infty}d\Delta_d \int_{-\infty}^{\infty}d\Delta_{}\frac{\exp\left(-2i\Delta_{}\Delta_d /\chi\right)}{\sqrt{\left(\gamma^{2}+(\Delta_{}+\Delta_d )^2\right)\left(\gamma^{2}+(\Delta_{}-\Delta_d )^2\right)}}
\nonumber \\
= &  \frac{2D^2\gamma}{\pi\chi}\int_{-\infty}^{\infty}d\Delta_d \int_{-\infty}^{\infty}d\Delta_{}\frac{\exp\left(-2i\Delta_{}\Delta_d /\chi\right)}{\sqrt{\left(\gamma^{2}+(\Delta_{}+\Delta_d )^2\right)\left(\gamma^{2}+(\Delta_{}-\Delta_d )^2\right)}}.
\end{align}
\end{widetext}
Then changing variables again, to $U=(\Delta_{}+\Delta_d )/\gamma$ and
$V=(\Delta_{}-\Delta_d )/\gamma$, we finally obtain
\begin{align}
\Gamma_{\infty} = &
\frac{D^2\gamma}{\pi\chi{}}\int_{-\infty}^{\infty}
\frac{\exp\left(\frac{-iU^2\gamma^2}{2\chi{}}\right)}{\sqrt{\left(U^2+1\right)}}dU
\int_{-\infty}^{\infty}\frac{\exp\left(
\frac{iV^2\gamma^2}{2\chi{}}\right)}{\sqrt{\left(V^2+1\right)}}dV \nonumber \\
= & \frac{D^2\gamma}{\pi\chi{}}
K_{0}\left(\frac{ i\gamma^2}{4\chi{}}\right)
K_{0}\left(\frac{-i\gamma^2}{4\chi{}}\right).
\label{final_Gamma_infinity}
\end{align}
Here $K_0$ is a modified Bessel function of zero order and
$\Gamma_{\infty}$ is necessarily real, since
$K_0(z^*)=K_0(z)^*$. Further, we note that
$\Gamma_{\infty}$ is a monotonically decreasing function of
$\chi$. We will later argue that this expression for
$\Gamma_{\infty}$ is valid for $\xi \gg 1$ in the strong coupling
case since the separation of the Lorentzians in equation
(\ref{ca_high_chirp1}), for which $\chi(t-t')\sim\gamma$, must
take place on a time-scale shorter than the time-scale for the
atomic dynamics. (This would be $1/D$ in the case of
strong-coupling and no chirp.) With such high chirp, resonant
modes which start to become occupied are moved far from resonance
before they can act back on the atom, and are replaced by empty
ones. In this way, the atom effectively sees an empty reservoir
and obeys Markovian dynamics. A solution for the behaviour of
$c_{a}(t)$ follows trivially:
\begin{align}
c_{a}(t) & \approx \exp\left(-\Gamma_{\infty}\;t/2\right).
\label{Markovian_decay}
\end{align}
Figure \ref{DecayRateFig} shows the agreement between this
approximate analytic result and a numerical simulation which used
a bath of discrete field modes and did not make the Markov
approximation. In fact, the result (\ref{Markovian_decay}) also
holds in the weak-coupling regime, independent of the rate
$\chi$, as discussed in sections \ref{results_weak} and
\ref{conclusions}. For this reason as $\chi\longrightarrow 0$
the static reservoir result $\Gamma_{\infty} = 2D^2/\gamma$,
equation (\ref{markov_const}) is recovered. However, in the
strong coupling case, this limit cannot be attained with equation
(\ref{eq:def_xi}), since, eventually, the constraint $\xi \gg 1$
is violated. In addition, we know that for $\chi=0$ we should
find the static decay result $\gamma$ implied by equation
(\ref{exact_ca_static}). Nevertheless, since $2D^2/\gamma >
\gamma$ in the strong coupling regime, it is straightforward to
have an \emph{enhanced} decay rate (i.e.\ $\Gamma_{\infty} >
\gamma$) if $\xi$ is greater than unity, but not too high. In
figure \ref{DecayRateFig} we can find this for the low chirp part
of the $D=2\gamma$ curve ($12 \lesssim \chi/\gamma^2 \lesssim 38$), since
only then is $\Gamma_{\infty} / \gamma > 1$ (with $\xi \gg 1$);
for the rest of the curve the decay is \emph{inhibited} relative
to $\chi=0$. However, if $ D/\gamma $ is even modestly increased
the region of enhanced decay is increased substantially. It is
interesting to note the equation (\ref{final_Gamma_infinity}) can
be re-written as
  \begin{equation}
     \Gamma_{\infty} / \gamma = 
      2(D/\gamma )^2 \cdot  \left[\frac{ 2 |
          K_0(i/x)|^2}{\pi x} \right] ,
    \label{eq:Gam_inf_rewrite}
  \end{equation}
where $x=4\chi/\gamma^2$. In this expression the inequality $2|
K_0(i/x)|^2/\pi x \le 1$ holds, with equality as $x$ (i.e.\
$\chi$) tends to zero. Since the factor $D/\gamma$ is always
greater than unity for strong coupling, we may have either
\emph{enhanced} or \emph{inhibited} Markovian decay, depending
on the value of $\chi$. Equation (\ref{eq:Gam_inf_rewrite})
can also be used to motivate the constraint $\xi \gg 1$
introduced above. Since the time-scale $t_{at}$ for atomic
dynamics is now $1/ \Gamma_{\infty}$, and the factor $2|
K_0(i/x)|^2/\pi x$ is always less than (or equal to) unity, it
follows that $t_{at} \ge \gamma/2D^2$. Since the time-width of
the kernel is approximately $\gamma/\chi$, the system will be
Markovian if $\gamma/\chi \ll t_{at}$. Since $\xi \sim
\pi\chi/D^2$ for $D>\gamma$, this will always be true if $\xi
\gg 2\pi$.

For very large chirp, compared to $\gamma^2$, we can expand the
Bessel functions in equation (\ref{final_Gamma_infinity}) to
obtain the approximate result
\begin{equation}
 \Gamma_{\infty} / \gamma  \longrightarrow 
  \left(\frac{D}{\gamma} \right)^2
  \cdot \frac{\gamma^2}{\pi\chi}
  \left[\ln\left(4\chi/\gamma^{2}\right)\right]^{2}
  ,
  \label{eq:Gamma_limit}
\end{equation}
which slowly decreases as $\chi/\gamma^{2}$ is increased.

\subsubsection{Intermediate Chirp Rates}
\label{results_strong_intermediate}

As already discussed in section \ref{static_reservoir_strong}, in
the static case the excitation of the reservoir modes moves
outwards from $\omega_{0}$ over several Rabi cycles towards final
peaks at $\omega_{0}\pm \Omega/2 $. Now in the
intermediate-chirp case, the frequency of the modes is increased
by approximately the same amount over one Rabi-cycle. That is,
the occupied modes are brought back into resonance with the atom
on every Rabi-cycle. This can have a profound effect on the
dynamics of the atomic excited state probability,
$\vert{}c_{a}(t)\vert^{2}$, since occupied modes with angular
frequency in the vicinity of $\omega_{0}$ have the strongest
interaction with the atom, as can be seen from equation
(\ref{chirp_dynamics1}).

\begin{figure}[] 
\includegraphics[width=0.8\columnwidth]{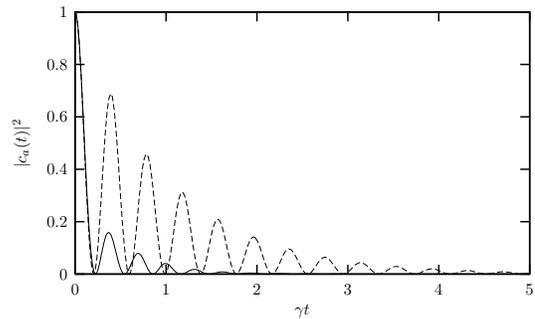}
\caption{Figure showing enhanced decay of the envelope of
  $\vert{}c_{a}(t)\vert^2$ caused by manipulation of the
  reservoir modes to which the atom is coupled (by a linear
  chirp).  We note that the Rabi period is modified by
  manipulation of the reservoir mode frequencies. Parameters are
  $D=8\gamma$ and for the solid line, $\chi=20\gamma^2$. Thus
  $\xi\approx1.0$. For comparison, the dashed line shows
  $\vert{}c_{a}(t)\vert^2$ for the case with no chirp
  ($\chi=0$).}
\label{enhanced_decay}
\end{figure}

\begin{figure}[] 
\includegraphics[width=0.8\columnwidth]{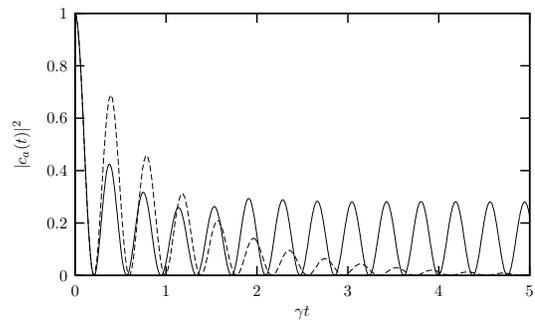}
\caption{Inhibition of decay and trapping of population with a
  chirped reservoir. The parameters are $D=8\gamma$ and for the
  solid line, $\chi=8.4\gamma^2$ (which implies
  $\xi\approx0.42$). For comparison, the dashed line shows
  $\vert{}c_{a}(t)\vert^2$ for the case with no chirp
  ($\chi=0$).}
\label{testfig}
\end{figure}

\begin{figure}[h] 
\includegraphics[width=\columnwidth]{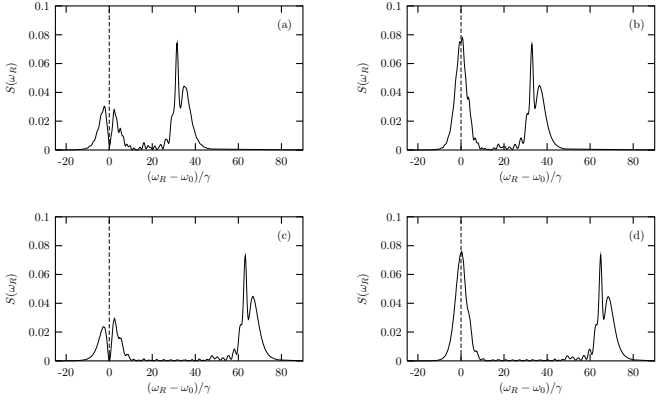}
\caption{ The reservoir excitation spectrum during the stable
  oscillations of figure \ref{testfig}. The horizontal axis
  shows the scaled time-dependent detuning from the atomic
  transition frequency (vertical dashed line), where
  $\omega_{R}=\omega+\chi{}t$. For (a-d) $\gamma t\sim$ 3.81,
  4.00, 7.60, 7.79, and other parameters are as for figure
  \ref{testfig}, i.e.\ $D=8\gamma, \chi=8.4\gamma^{2}$. The times
  selected correspond to approximately: (a) 10; (b) 10.5; (c) 20;
  and (d) 20.5 Rabi oscillations. }
\label{fig:chirp_bath_spectrum}
\end{figure}

In this regime of intermediate chirp rate ($\xi\sim1$) and strong
coupling ($D\gg\gamma$), we must rely on numerical results. As
illustrated in figure \ref{landscape_fig}, two free parameters,
i.e.\ $D/\gamma$ and the rate of redistribution of reservoir mode
frequencies, label the dynamics of the system. A numerical
search was conducted over this plane, and it was found that Rabi
oscillations were inhibited over most of this region. The
explanation for this is that occupied modes of the reservoir are
moved from resonance before they have a chance to re-populate the
atom. A typical example is shown in figure \ref{enhanced_decay}.
Additionally, for $D/\gamma\lesssim15$, there is a particular
chirp-rate for which a `recycling condition' is fulfilled and the
vacuum Rabi-splitting in the reservoir doesn't occur. A
significant population may become trapped in a cycle of resonant
exchange between atom and reservoir, and Rabi oscillations may
persist for much \emph{longer} than in the static case. The
atomic behaviour is shown in figure \ref{testfig}. We also note
that the oscillation frequency increases with increasing
chirp-rate as is seen empirically from simulations. In figure
\ref{fig:chirp_bath_spectrum} we see the corresponding spectrum
of excitation in the reservoir for two pairs of times. In each
case the structured peak at higher frequencies is sufficiently
detuned for this reservoir population to be effectively decoupled
from the atomic dynamics. Therefore, excitation simply moves to
higher frequencies at later times, as may be seen by comparing
the spectra after $10$ and $20$ Rabi cycles. This non-interacting
peak has an area of approximately a half, which represents the
population that was `pushed' to frequencies above $\omega_{0}$
during the Rabi splitting at early times. We see that the left
hand peak, which is centred at the atomic transition, remains on
resonance as individual mode frequencies pass through this
region. It does invert periodically during the time evolution,
but this population exchange, or recycling, between the atom and
\emph{locally resonant} reservoir modes is very stable. For
example, we can see that at the same part of the cycle the shape
of this peak is relatively unchanged between (a) and (c) [and
between (b) and (d)] in figure \ref{fig:chirp_bath_spectrum}.
This behaviour is quite different to that seen for the static
reservoir in figure \ref{intermediate_chirp_fig}.

\subsubsection{Low Chirp Rates}
\label{results_strong_low}

\begin{figure}[!h] 
\includegraphics[width=0.8\columnwidth]{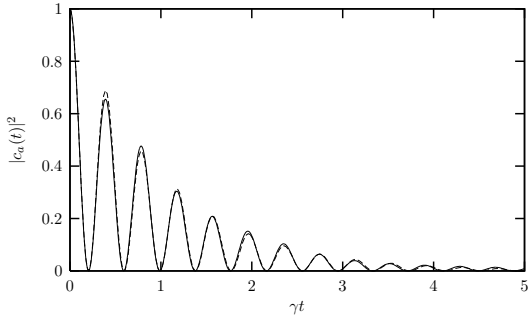}
\caption{For low chirp-rates, the atom experiences coupling to
  the same set of modes for several Rabi cycles and,
  consequently, the effects of the chirp are small. Parameters
  chosen were $D=8\gamma$, and for the solid line,
  $\chi=2\gamma^2$, which corresponds to $\xi\approx0.1$. For
  comparison, the dashed line shows $\vert{}c_{a}(t)\vert^2$ for
  the case with no chirp ($\chi=0$).}
\label{high_Q_low_chirp_fig}
\end{figure}

In the regime of strong coupling and very low chirp ($\xi\ll1$)
the basic behaviour of the system will be to perform Rabi
oscillations of the type found in equation
(\ref{exact_ca_static}). The weak chirp will be expected to
perturb the dynamics only slightly. In the example shown in
figure \ref{high_Q_low_chirp_fig} the difference between the
chirped and unchirped dynamics is very small, although it may be
noticed the the chirped bath results in a slightly raised Rabi
frequency. This frequency shift might be observable if the decay
rate is low enough that a large phase shift accumulates in the
Rabi oscillations. To evaluate the size of this effect we examine
a perturbative solution found by expanding the denominator of
equation (\ref{eq:integro_diff3}) in powers of $\chi$. Thus, for
the kernel of equation (\ref{eq:integro_diff3})
\begin{widetext}
\begin{align}
   K(\tau) = & \frac{D^2\gamma}{\pi}
  \int_{-\infty}^{\infty}d\Delta
\frac{ \exp(-i\Delta\tau) 
}{
\sqrt{
\left[ \gamma^2 + (\Delta + \chi\tau/2)^2 \right]
\left[ \gamma^2 + (\Delta - \chi\tau/2)^2 \right]
}}
\nonumber\\
 \sim& \frac{D^2\gamma}{\pi}
  \int_{-\infty}^{\infty}d\Delta
  \frac{e^{-i\Delta\tau}
  }{\gamma^2+\Delta^2}  \left[
  1 -
  \frac{\chi^2\tau^2}{4} \frac{\gamma^2-\Delta^2}
  {(\gamma^2+\Delta^2)^2} 
  + ...
  \right],
  \label{eq:low_chi_kernel}
\end{align}
\end{widetext}
to order $\chi^2$. The method of pseudomodes (see section
\ref{static_reservoir} and Ref.\ \cite{garraway1997}) could be
used to analyse this kind of kernel structure, but here we will
simply examine the Laplace transform of equation
(\ref{eq:integro_diff3}) to find the perturbation of the Rabi
frequency. To do this we first perform the integral over the
detuning $\Delta$ in equation (\ref{eq:low_chi_kernel}) so that
with the approximate kernel (and given $\tau>0$)
\begin{equation}
  K(\tau) = D^2 e^{-\gamma\tau} \left[ 1 -
  \frac{\chi^2\tau^2}{16\gamma^2}
  \left( 1 + \gamma\tau + (\gamma\tau)^2  \right)
  \right]
  .
  \label{eq:low_chi_dint}
\end{equation}
If we Laplace transform equation (\ref{eq:integro_diff3}) and
solve for $\tilde{c}_a(s)$, the transform of $c_a(t)$, we find
\begin{equation}
  \tilde{c}_a(s) = \frac{1}{s+ \widetilde{K}(s)}
  ,
  \label{eq:LTdiff3}
\end{equation}
where $\widetilde{K}(s)$ is the transform of equation
(\ref{eq:low_chi_kernel}) and we let $c_a(0)=1$. Using the
approximate kernel (\ref{eq:low_chi_dint}) we will find that
\begin{eqnarray}
  \widetilde{K}(s) & = & 
 \frac{D^{2}}{s+\gamma}\bigg[1
 -\frac{\chi^{2}}{8\gamma^{2}(s+\gamma)^{2}}
 \nonumber \\
& & \hspace{7mm}-\frac{3\chi^{2}}{8\gamma(s+\gamma)^{3}}-\frac{3\chi^{2}}{2(s+\gamma)^{4}}
 \bigg]
  .
  \label{eq:LTK}
\end{eqnarray}
If we momentarily set $\chi$ to zero we see that, if we
substitute equation (\ref{eq:LTK}) into equation
(\ref{eq:LTdiff3}), the poles in the right hand side of equation
(\ref{eq:LTdiff3}) are located at $s= -\gamma/2 \pm i\Omega/2$,
where $\Omega$ is given in equation (\ref{Omega_definition}).
This represents the damped oscillations of equation
(\ref{exact_ca_static}). If we now admit a finite value of
$\chi$, we seek the perturbed locations of the poles in the
approximate form
\begin{equation}
   s=  -\frac{\gamma}{2} \pm i\frac{\Omega}{2} + \delta
   \chi^2 .
  \label{eq:LTpoles}
\end{equation}
The factor $\delta$ is found by substitution into the equation $
s+ \widetilde{K}(s) =0$. Since the modification of the decay rate
is small, and not of interest, we examine the imaginary part of
$\delta$ to find that Im$(\delta) \sim \pm 1/(8\Omega\gamma^2)$
and then the new Rabi frequency $\Omega'$ is approximately
\begin{equation}
   \Omega' \sim \Omega \left( 1 +
     \frac{\chi^2}{4\Omega^2\gamma^2} \right)
   .
  \label{eq:LTNewOmega}
\end{equation}
In the above we also used the strong coupling approximation
$(D, \Omega) \gg\gamma$. 
Even if the correction to the Rabi frequency is small, the effect
may be noticeable since there will be an accumulated phase shift
over time. Since the largest usable time in this limit is of
order $1/\gamma$, it follows that the largest phase difference is
approximately given by
\begin{equation}
  \Delta\Phi  = (\Omega' - \Omega)T \sim
       \left(
     \frac{\chi^2}{4\Omega^2\gamma^2} \right) 
 \left(\frac{\Omega}{\gamma}\right)
       .
  \label{eq:low_chirp_phase_diff_max}
\end{equation}
Thus even if the first term in brackets is small, since it is the
correction to the Rabi frequency in equation
(\ref{eq:LTNewOmega}), the strong coupling regime ensures the
second term is large and so the resulting phase difference may be
of a reasonable size.  Note that the effect will be made more
visible, not only by increasing the chirp $\chi$, but also by
\emph{reducing} the strong atom-cavity coupling.

\subsection{Weak Coupling}
\label{results_weak}
In the limit $D\ll\gamma$, the behaviour of
$\vert{}c_{a}(t)\vert^{2}$ is Markovian. This remains true,
irrespective of how the reservoir mode frequencies are
manipulated. Therefore, for weakly-coupled reservoirs, the
analysis of section \ref{results_strong_high}, culminating in
equation (\ref{Markovian_decay}), is valid for all chirp-rates.
The region of weak coupling corresponds to the horizontal strip
at the bottom of figure \ref{landscape_fig}. Note that the
expansion \cite{gradshteyn2000}
\begin{equation}
\lim_{\vert{}z\vert\rightarrow\infty}\left\{K_{0}(z)\right\} \sim
\sqrt{\frac{\pi}{2z}}e^{-z}\left(1
-\frac{1}{8z}+\frac{9}{128z^2}
+\ldots\right)
\end{equation}
implies that the decay constant $\Gamma_{\infty}$ tends to the
result of the static reservoir, equation (\ref{markov_const}), in
the low-chirp limit, as it should. We can also find the leading
correction, so that as $\chi \longrightarrow 0$
\begin{equation}
\Gamma_{\infty}  \longrightarrow
\frac{2D^2}{\gamma}
\left( 1 - \frac{ 2 \chi^2}{\gamma^4} + \ldots  \right)
.
\end{equation}

\section{Conclusion}
\label{conclusions}
\label{observable_effects}

We have investigated the effects on the dynamics of an open
quantum system of a particular manipulation of the reservoir mode
frequencies. While other authors have discussed reservoirs with
time-dependent properties, we have here considered the subtle
case where all \emph{macroscopic} properties of the reservoir
remain \emph{static}, while the microscopic structure of the
reservoir changes. This problem is intriguing, since the effects
are non-trivial, and cannot easily be guessed in advance. It is
tempting to think that since the only difference with the usual,
static case is a redistribution of reservoir mode
\emph{populations}, only non-Markovian systems would be affected.
This is not the case, however, and an effect can be seen for both
Markovian and non-Markovian systems.

We have characterised the nature of the dynamics in three
different regions of parameter space (see figure
\ref{landscape_fig}) and found conditions for each type of
behaviour in section \ref{results}. It is shown that manipulation
of the reservoir mode frequencies may alter the \emph{nature} of
the dynamics (from non-Markovian to Markovian behaviour) when the
chirp-rate is high ($\xi\gg1$). Analytical results are
presented for the decay in this case (equations
(\ref{final_Gamma_infinity}) and (\ref{Markovian_decay})). The
resulting decay may be enhanced or inhibited, compared to the
un-chirped case. In the case of strong coupling and low chirp,
the chirp manifests itself as a frequency shift in the Rabi
oscillations (equation (\ref{eq:LTNewOmega})).

In order to realise the effects of section \ref{results} in the
context of optical cavity QED it would be necessary to construct
a high-Q cavity and a reservoir whose mode frequencies could be
manipulated as described above, i.e.\ without altering the
envelope of the coupling constants near the atomic resonance.
Ideally, if the resonant frequency of the cavity is to remain
unchanged (at $\omega_{0}/2\pi$), its cavity
mirrors must be fixed, too. Since there will be losses through
the cavity mirrors, i.e.\ a coupling to the external environment,
the bath mode frequencies may be altered by changing the
properties of that external environment. This would be
sufficient to isolate the basic effect from, for example,
changing the position of the cavity resonance (as in Ref.\
\cite{Law95}). The most simple way of doing this would be to
have an inner and outer cavity, with lengths $\ell$ and $L$. We
can suppose that the outer mirror can be moved by a piezoelectric
actuator, or similar device, which will change the length $L$.
Then for small fractional changes in the outer cavity length we
can write
\begin{equation}
\frac{\partial\omega_{}}{\partial{}t}\approx-\frac{\omega_{}}{ L }\frac{\partial{} L }{\partial{}t}\approx-\frac{\omega_{0}}{ L }\frac{\partial{} L }{\partial{}t},
\label{moving_mirror}
\end{equation}
so that
\begin{equation}
\chi\approx-\frac{\omega_{0}}{ L }\frac{\partial{} L }{\partial{}t}.
\end{equation}
To illustrate the potential size of any observable effects, we
consider some specific cases. First we take an optical atomic
transition with angular frequency
\mbox{$\omega_{0}/2\pi\sim3.5\times10^{14}$ Hz} and a 40 $\mu$m inner
cavity for which \mbox{$\gamma/2\pi\sim4.1$ MHz} and \mbox{$D/2\pi\sim 34$ MHz}
\cite{boca2004,miller2005}. This is in the strong coupling
regime. For the outer cavity we let $L=$ 1cm, and we assume here
that the inner cavity is diffractively coupled to the outer
cavity to enhance the mode density, whilst maintaining a strong
directional property. If we can move the out-most mirror at a
speed of 0.1 ms$^{-1}$ for the short time necessary, we find that
\mbox{$\chi \sim 2.2\times10^{16}$ s$^{-2}$}. Then $\xi \approx 1.52$,
which is enough to enter the intermediate chirp region indicated
in figure \ref{landscape_fig} and discussed in section
\ref{results_strong_intermediate}. If we would increase the
mirror speed to 0.65 ms$^{-1}$, $\xi$ increases to approximately
ten, which places us in the high chirp, strong coupling regime of
figure \ref{landscape_fig}. In this case equation
(\ref{final_Gamma_infinity}) applies and we obtain an enhancement
of decay: $\Gamma_{\infty}/\gamma = 5.05$. This enhanced decay
can be increased even further if, for example, the cavity decay
rate is reduced by a factor of ten: in that case
$\Gamma_{\infty}/\gamma = 13.6$. As an example of inhibited decay
we can simply reduce $D$ so that $D\ll\gamma$. In this case we
are in the weak coupling regime and the inhibition of decay
relative to the static reservoir case is independent of $D$ and
given by a factor of approximately
\begin{equation}
 \frac{\Gamma_{\infty}}{\left(2D^2/\gamma\right)} 
     \approx  9.92 \times 10^{-4},
\end{equation}
when $\gamma/2\pi\sim 0.41$ MHz and the speed is 0.65 ms$^{-1}$,
as above.

We have seen that many of the effects described here could be
tested with currently available experimental parameters. For an
atom weakly coupled to a high-Q cavity, the inhibition of decay
and decoherence is considerable, with the possibility of
suppressing the decay rate by a factor of $10^{3}$. In the case
of strong coupling, atomic decay can be inhibited (for high chirp
$\chi$) or enhanced. The examples given show an enhancement of
decay by a factor of 5 or 14. This factor can be increased if the
ratio $D/\gamma$ is increased along with the smallest cavity
chirp consistent with $\xi \gg 1$. An attractive feature of this
model is that the suppression, or enhancement, could be turned on
and off at will by moving or not moving the mirror of an
\emph{outer} cavity. In the case of enhanced decay, this means
that the controlled motion of an outer cavity mirror can extract
energy from the inner cavity-atom system.

  This work was supported in part by the UK Engineering and
  Physical Sciences Research Council.


\begin{thebibliography}{99}
\bibitem{note1} After a time $t$ each mode increases its frequency by
$\chi{}t$, see (\ref{frequency_definition}).
\end{thebibliography}
\end{document}